\documentclass[cameraready]{Interspeech}
\usepackage{multirow}
\title{LLM-HB: Language-Aware LLM-Guided Hotword Biasing for Code-Switching ASR}

\author[affiliation={1,2,3}, orcid=0009-0004-5625-6184]{Yuxuan}{He}
\author[affiliation={2}, orcid=0000-0002-5813-9430]{Genshun}{Wan}
\author[affiliation={2}]{Pengcheng}{Li}
\author[affiliation={3}, orcid=0000-0002-6825-7473]{Gongping}{Huang}
\author[affiliation={2}, orcid=0000-0001-5575-4940]{Jianqing}{Gao}
\author[affiliation={1}]{Yanmin}{Qian}

\address{
    \mbox{$^1$ Auditory Cognition and Computational Acoustics Lab, Shanghai Jiao Tong University, Shanghai, China}
    \mbox{$^2$ AI research Institute, iFLYTEK Company Limited, Hefei, Anhui, China}
    \mbox{$^3$ School of Electronic Information, Wuhan University, Wuhan, Hubei, China}
}

\email{yuxuanhe@whu.edu.cn,gswan@iflytek.com,yanminqian@sjtu.edu.cn}


\usepackage{comment}


\begin{document}

\maketitle

\begin{abstract}
\vspace{-2mm}
Code-switching (CS) speech, characterized by language alternation within a single utterance, is prevalent in multilingual communities and poses challenges to automatic speech recognition (ASR) due to rapid language transitions. We propose LLM-HB, a language-aware large language model-guided hotword biasing framework for code-switching ASR (CS-ASR). The framework incorporates a mixture-of-experts (MoE) adaptor and an auxiliary language prediction head to better capture and differentiate multilingual representations, while hotwords are injected through LLM-based prompting. Experiments on the ASRU2019 code-switching challenge test set demonstrate that LLM-HB achieves a 20.30\% relative reduction in mixed error rate (MER) over the baseline, reaching 5.85\% MER under a 15-distractor hotword setting. To the best of our knowledge, this is the first work to integrate LLM-guided biasing into code-switching ASR. We further release the extracted ASRU2019 hotword list to support future research.
\end{abstract}

\vspace{-2mm}

\section{Introduction}

Code-switching (CS) refers to the alternation between multiple languages within a single utterance or discourse. Compared with monolingual settings, automatic speech recognition (ASR) faces additional challenges in CS scenarios due to frequent inter- and intra-sentence language transitions~\cite{Povey_ASRU2011_2011,gulati20_interspeech,8068205}. Although conventional ASR systems can be applied to code-switching speech similarly to monolingual data, early studies typically perform language identification prior to recognition or jointly model language identification and transcription~\cite{6289015,liu21d_interspeech}. More recent CS-ASR approaches mitigate cross-language confusion by incorporating language-aware components into the ASR architecture.

Prior work has explored various strategies to address cross-language interference in CS-ASR. A representative method employs a Transformer-based bi-encoder architecture that decouples English and Mandarin modeling using independently pre-trained encoders~\cite{lu20f_interspeech,NIPS2017_3f5ee243}. Beyond explicit structural separation, language-aware frameworks dynamically specialize representations for different languages~\cite{9747537,tian22c_interspeech}. Mixture-of-Experts (MoE) adaptors enable dynamic expert specialization to enhance language-dependent representation learning and distinguish language boundaries~\cite{10389798,DBLP:conf/iclr/ShazeerMMDLHD17,20260056940}. Other approaches incorporate frame- or token-level language posteriors into acoustic representations prior to decoding~\cite{10095878,10448335}, while cross-modality learning aligns speech and generated code-switching text embeddings in a shared latent space to better model switching dynamics~\cite{10096317}.

In contextual biasing, extensive efforts have focused on improving recognition of named entities and rare words. Prior to large language models (LLMs), contextual biasing methods typically integrated hotword embeddings with acoustic representations during decoding. Such integration can be performed within the speech encoder~\cite{shakeel24_interspeech,sudo25_interspeech}, after the encoder~\cite{9415054}, or before the ASR decoder~\cite{10446106}. With the introduction of LLMs, hotwords or their embeddings are incorporated as prompts together with acoustic representations. One line of work generates an initial transcription and refines it by retrieving semantically similar hotwords~\cite{10832154}. Another adopts contrastive learning to align acoustic and hotword embedding spaces, followed by nearest-neighbor retrieval (e.g., FAISS~\cite{8733051}) for LLM conditioning~\cite{gong25_interspeech,kong25_interspeech}.


Despite these advances, most CS-ASR approaches focus on mitigating cross-language interference within acoustic and linguistic modeling without explicitly considering contextual biasing. Conversely, existing contextual biasing techniques are primarily developed for monolingual ASR systems. To date, limited efforts have attempted to integrate contextual biasing into CS-ASR, and these approaches do not fully exploit the capability of LLMs for adaptive multilingual biasing~\cite{wan2024xcbeffectivecontextualbiasing}.

\begin{figure*}[t!]
	\centering
	\includegraphics[width=\linewidth]{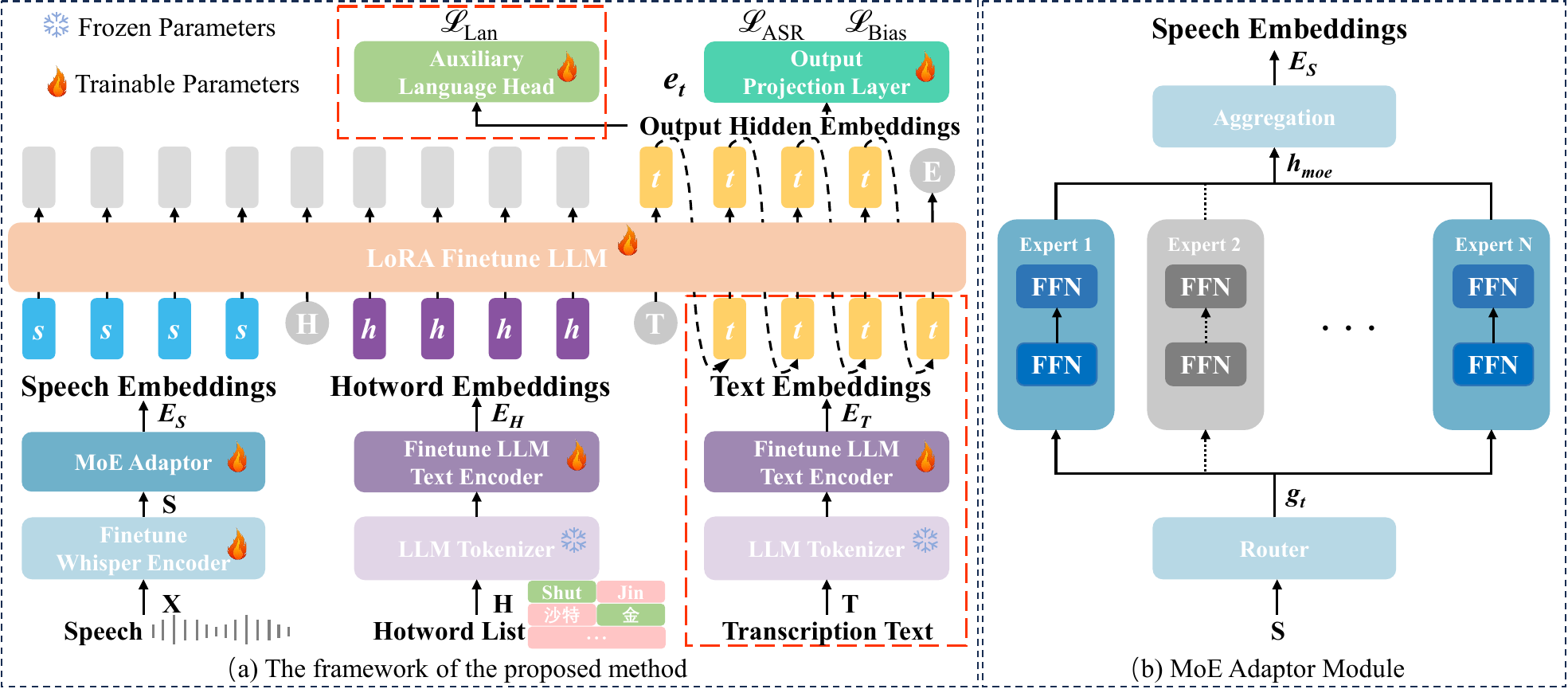}
    \vspace{-12pt}
    \caption{The framework of the proposed large language model-guided hotword biasing method and MoE adaptor.}
	\label{fig1}
    \vspace{-13pt}
\end{figure*}


To address these limitations, we propose \textbf{LLM-HB}: a language-aware \textbf{L}arge \textbf{L}anguage \textbf{M}odel\textbf{-}guided \textbf{H}otword \textbf{B}iasing method for CS-ASR. Our approach integrates MoE adaptors after the speech encoder to enable language-specialized acoustic representation learning. An auxiliary language prediction head is attached to the LLM representations and jointly optimized with the ASR objective to provide explicit language supervision, enhancing language discrimination. Hotwords are incorporated through LLM prompting, enabling adaptive contextual biasing under multilingual settings. By jointly modeling language separation and LLM-guided biasing, the proposed framework reduces cross-language interference while improving recognition of rare and named entities in code-switching speech.

We evaluate LLM-HB on the ASRU2019 code-switching challenge test set~\cite{shi2020asru2019mandarinenglishcodeswitching} under a 15-distractor hotword setting\footnote{Hotword lists are available at: \url{https://anonymous.4open.science/r/IS26-ASRU2019-Hotword-List/}}, achieving a 20.30\% relative reduction in mixed error rate (MER) over the baseline and reaching 5.85\% MER. In addition, LLM-HB achieves 4.94\% character error rate (CER) and 13.21\% word error rate (WER), with CER and WER reduced by 1.98\% and 49.29\% relative to the baseline, respectively.


\vspace{-2mm}

\section{Methodology}

In this section, we present the overall architecture of LLM-HB, including the MoE adaptor, the auxiliary language head, and the training objectives, as illustrated in Figure~\ref{fig1}.

\subsection{Overall Architecture}

Figure~\ref{fig1}(a) presents the overall architecture of LLM-HB. 
Given an input speech spectrogram $X \in \mathbb{R}^{M \times T}$, where $M$ denotes the number of Mel-frequency bins and $T$ the number of time frames. Acoustic features $S$ are extracted using a pre-trained Whisper encoder~\cite{pmlr-v202-radford23a} and fine-tuned during training. 
The encoded features are passed through an MoE adaptor to produce language-specialized speech embeddings $E_S \in \mathbb{R}^{T \times D}$, where $D$ is aligned with the LLM representation dimension.

The hotword list $H = {h_0, h_1, \ldots, h_N}$, containing $N$ hotwords and a special $<\text{no-bias}>$ token, is tokenized using a frozen LLM tokenizer.
Let $N'$ denote the total number of tokens.
The tokenized sequence is encoded by the LLM text encoder to obtain hotword embeddings $E_H \in \mathbb{R}^{N' \times D}$.
During training, the ground-truth transcription is processed similarly to produce target text embeddings $E_T \in \mathbb{R}^{L \times D}$, where $L$ denotes the number of transcription tokens.

The speech embeddings $E_S$ and hotword embeddings $E_H$ are fed into a LoRA-adapted~\cite{DBLP:conf/iclr/HuSWALWWC22} LLM. 
Special marker tokens \textcircled{H}, \textcircled{T}, and \textcircled{E} denote the beginning of hotword embeddings, transcription embeddings, and the end-of-sequence token, respectively. 
During training, the LLM input sequence is constructed as 
$[E_S, \textcircled{H}, E_H, \textcircled{T}, E_T, \textcircled{E}]$.

The LLM outputs hidden states $e_t$, which are projected to the vocabulary space for token prediction. 
An auxiliary language prediction head is attached to $e_t$ to provide explicit language supervision. 
The modules enclosed by the red dashed boxes in Figure~\ref{fig1}(a), including the auxiliary language head and the ground-truth text encoding branch, are used only during training and removed during inference.

\subsection{Language-Specialized MoE Adaptation and Auxiliary Supervision}








As illustrated in Figure~\ref{fig1}(b), the MoE adaptor performs dynamic expert selection and aggregation to produce language-specialized speech embeddings. 
Given acoustic features $S$, a router computes top-$K$ gating weights, and the selected experts process the input independently. 
Their weighted outputs are aggregated into an intermediate representation $h_{\text{moe}}$, defined as

\begin{equation}
g_t = \text{Softmax}(\text{Top-K}(\text{Router}(S), k))
\end{equation}
\begin{equation}
h_{\text{moe}} = \sum_{i \in \mathcal{K}} g_{t,i} \, \text{FFN}_i(S)
\end{equation}

The aggregated representation $h_{\text{moe}}$ is further transformed by a feed-forward layer followed by layer normalization (the \emph{Aggregation} block in Figure~\ref{fig1}(b)) to obtain the final speech embeddings $E_S$:

\begin{equation}
E_S = \text{LayerNorm}(\text{FFN}(h_{\text{moe}}))
\end{equation}

To provide explicit language supervision, an auxiliary language prediction head is attached to the LLM hidden states $e_t$. 
The language posterior is computed as

\begin{equation}
P_{l_t} = \text{Softmax}(\text{LayerNorm}(\text{FFN}(e_t)))
\end{equation}

where $P_{l_t}$ denotes the predicted language distribution for the $t$-th token. 
This auxiliary objective encourages language-aware representations and stabilizes expert specialization.

\subsection{LLM-Guided Hotword Biasing}

To encourage hotword recognition, we introduce an LLM-guided biasing objective.
Given acoustic input $X$ and a hotword set $H$, the model is optimized to increase the posterior probability of hotword tokens during decoding.

The biasing loss is defined as
\begin{equation}
\mathcal{L}_{Bias} = -\sum_{t \in \mathcal{H}} \log p(y_t|X,H)
\end{equation}
where $y_t$ denotes the ground-truth token at position $t$, and $\mathcal{H}$ represents the set of token positions corresponding to hotwords in the target sequence. 
Maximizing this objective encourages the model to prioritize bias-relevant tokens while preserving standard modeling of non-hotword regions.

\subsection{Loss Functions}

The overall training objective jointly optimizes the ASR loss, the hotword biasing loss, and the auxiliary language supervision loss within a unified end-to-end framework:

\begin{equation}
\mathcal{L} = \mathcal{L}_{ASR} + \lambda_1\mathcal{L}_{Bias} + \lambda_2\mathcal{L}_{Lan}
\label{eq6}
\end{equation}

The ASR loss $\mathcal{L}_{ASR}$ is defined as the standard cross-entropy (CE) loss over the target transcription tokens. 
The auxiliary language loss $\mathcal{L}_{Lan}$ is also formulated as a CE loss over the predicted language distribution at each output token. 
Combined with the biasing objective $\mathcal{L}_{Bias}$, the model is trained end-to-end to balance transcription accuracy, language discrimination, and hotword emphasis.








\section{Experiment}

\subsection{Dataset and Experimental Setup}

Experiments were conducted on three benchmark datasets: the 1000-hour Mandarin corpus AISHELL-2~\cite{du2018aishell2transformingmandarinasr}, the 960-hour English corpus LibriSpeech~\cite{7178964}, and the 200-hour code-switching dataset ASRU2019-CS.
Depending on the training configuration, models were trained either on the full multi-dataset setup (AISHELL-2, LibriSpeech, and ASRU2019-CS) or solely on ASRU2019-CS, while evaluation was consistently performed on ASRU2019-CS. To construct the evaluation hotword lists, all English words in the test set were selected, and Chinese named entities were extracted using HanLP\footnote{\url{https://github.com/hankcs/HanLP}}~\cite{he-choi-2021-stem}
. These words were used to form hotword lists containing 15 distractors per utterance for evaluation.

We report the mixed error rate (MER), which combines word error rate (WER) for English and character error rate (CER) for Mandarin, together with bias-oriented metrics, including B-MER, B-WER, and B-CER~\cite{le21_interspeech}.

For the proposed LLM-HB framework, hyperparameters were set as follows. We employed Whisper-medium\footnote{\url{https://github.com/openai/whisper}}
 as the speech encoder for its strong multilingual capability in code-switching scenarios. As the LLM backbone, we adopted Qwen3-4B\footnote{\url{https://huggingface.co/Qwen/Qwen3-4B}}
, which provides a favorable trade-off between efficiency and representational capacity. The Whisper encoder produces 1024-dimensional acoustic embeddings, which are projected to 2560 dimensions via a dense MoE adaptor with two experts and top-2 gating to match the LLM input space.


For hotword biasing, the predefined hotword list is tokenized and encoded by the frozen LLM text encoder to obtain hotword embeddings. During training, the ground-truth transcription is processed in the same way to ensure consistency in the shared embedding space. The auxiliary language head takes the 2560-dimensional LLM outputs and applies an FFN to project them into three dimensions corresponding to Mandarin, English, and other tokens.

\begin{table*}[t!]
    \caption{Performance comparison of the proposed LLM-HB method and its variants across different datasets. Base model refers to the Whisper medium encoder with a linear projection layer and a frozen Qwen3-4B LLM. All metrics are reported in percentage (\%).}
    \label{tabel1}
    \vspace{-6pt}
    \centering
    \renewcommand{\arraystretch}{0.9}
    \begin{small}
    \begin{tabular*}{\linewidth}{@{\extracolsep{\fill}} l c c c c c c c }
        \toprule
        \small Method & \small Training Dataset & \small MER & \small B-MER & \small CER & \small B-CER & \small WER & \small B-WER \\
        \midrule

        \small Base Model & \multirow{5}{*}{\parbox{2.5cm}{\centering ASRU2019-CS-Train-Set}} & 9.59 & 26.91 & 7.03 & 25.45 & 30.44 & 27.27 \\
        \small + MoE Adaptor & & 9.65 & 27.18 & 7.06 & 25.57 & 30.73 & 27.58 \\
        \small + Language Head & & 9.30 & 26.27 & 6.77 & 24.50 & 29.87 & 26.72 \\
        \small + Hotword Prompt & & 7.48 & 16.00 & 5.92 & 10.49 & 20.21 & 17.75 \\
        \small + MoE + LH + HP (LLM-HB) & & 7.36 & 14.14 & 6.02 & 8.11 & 18.29 & 16.07 \\
        \midrule
        \small Base Model (baseline) & \multirow{5}{*}{\parbox{3cm}{\centering ASRU2019-CS-\\Train-Set \\ + Librispeech \\+ Aishell-2}} & 7.34 & 22.11 & 5.04 & 19.26 & 26.05 & 22.90 \\
        \small + MoE Adaptor & & 7.19 & 21.84 & \textbf{4.89} & 18.51 & 25.94 & 22.80 \\
        \small + Language Head & & 7.21 & 21.96 & 4.91 & 18.79 & 25.94 & 22.89 \\
        \small + Hotword Prompt & & 6.41 & 9.52 & 5.47 & 3.03 & 14.07 & 11.65 \\
        \small \textbf{+ MoE + LH + HP (LLM-HB)} & & \textbf{5.85} & \textbf{8.99} & 4.94 & \textbf{3.01} & \textbf{13.21} & \textbf{10.93} \\
        \bottomrule
    \end{tabular*}
    \end{small}
    \vspace{-12pt}
\end{table*}

During training, we optimized the model using AdamW with a triangular learning rate schedule. One cycle spanned 47,040 steps, and the model was trained for two cycles (94,080 steps). The peak learning rate was $1.5 \times 10^{-4}$.
In each mini-batch, each utterance had a 50\% probability of being selected for biasing, from which 1–4 tokens were randomly sampled to form the hotword list.
Training was conducted on approximately 2,200 hours of speech using 8 MLU590 GPUs. The loss weights in Equation~\ref{eq6} were set to $\lambda_1 = 0.6$ and $\lambda_2 = 0.3$.


\subsection{Main Results of LLM-HB Method}

Table~\ref{tabel1} presents the performance of the proposed LLM-HB method and its variants on the ASRU2019-CS test set under two training settings: (1) training solely on the ASRU2019-CS training set (200 hours), and (2) joint training on all three datasets (ASRU2019-CS, LibriSpeech, and AISHELL-2), totaling approximately 2,200 hours. Owing to differences in training data scale and model capacity, direct comparison with prior work is omitted to ensure a fair evaluation setting.

When trained only on the 200-hour code-switching dataset, the proposed modules exhibit distinct behaviors. The MoE adaptor alone leads to a slight performance degradation (MER increases from 9.59\% to 9.65\%), suggesting that the additional parameters introduced by the adaptor may require larger-scale data for effective optimization. In contrast, both the auxiliary language head and the hotword prompt yield consistent improvements, achieving MERs of 9.30\% and 7.48\%, respectively. When all three components are integrated, the complete LLM-HB further improves MER to 7.36\%, demonstrating that the proposed framework remains effective even under limited training data. This suggests that, although the MoE adaptor alone may be difficult to optimize effectively under limited data, its interaction with language supervision and biasing signals still provides complementary representational benefits when trained jointly. Notably, the hotword prompt delivers the most significant individual gain, reducing MER to 7.48\% and substantially lowering B-MER from 26.91\% to 16.00\%, confirming its strong biasing capability in low-resource conditions.

Under larger-scale multi-dataset training (2,200 hours), all modules become consistently beneficial. The complete LLM-HB model, integrating all three components, achieves the best MER of 5.85\%, corresponding to a 20.30\% relative reduction over the baseline (7.34\%). Notably, the MoE adaptor now contributes positively, indicating that sufficient data enables expert specialization to emerge more reliably.

The improvements are particularly pronounced on hotword-related metrics, with B-MER decreasing from 22.11\% to 8.99\%, B-CER from 19.26\% to 3.01\%, and B-WER from 22.90\% to 10.93\%. Importantly, the auxiliary language head also plays a structural role in shaping MoE behavior. Although the routing mechanism is not explicitly supervised, the language prediction loss is applied to the shared LLM representations obtained after expert aggregation. This supervision encourages the fused embeddings to retain language-discriminative cues, which in turn drives the gating network to allocate tokens to experts in a more language-consistent manner through backpropagation. Furthermore, since the injected hotwords span both Mandarin and English, the biasing objective interacts with the language-aware supervision, reinforcing expert specialization toward language-aligned patterns. As a result, expert differentiation emerges implicitly, enabling more effective utilization of bilingual hotword information.

These results indicate that the three modules operate complementarily: the MoE adaptor provides capacity for specialization, the auxiliary language head guides language-aware representation learning and indirectly influences expert routing, and the hotword prompt introduces targeted biasing signals.

Comparing the two training setups, all methods benefit substantially from larger-scale training. The relative improvements remain consistent across variants, with the complete LLM-HB achieving a MER of 5.85\% under the full multi-dataset setting. Under the code-switching-only setting, the complete LLM-HB achieves 7.36\% MER, indicating that the framework remains effective even with 200 hours of training data, while larger-scale training further unlocks the benefits of expert specialization.

Notably, the gap between hotword performance and overall performance (B-MER vs. MER) narrows under larger-scale training, suggesting improved generalization to both in-vocabulary and out-of-vocabulary terms.

\begin{table}[t!]
    \caption{Ablation on expert configurations. All metrics are reported in percentage (\%).}
    \label{tabel2}
    \vspace{-6pt}
    \centering
    \renewcommand{\arraystretch}{0.9}
    \begin{small}
    \begin{tabular*}{\linewidth}{@{\extracolsep{\fill}} l c c c c c c c }
        \toprule
        \small Configuration & \small MER & \small CER & \small WER \\
        \midrule
        \small 4 choose 4 & 7.33 & 5.01 & 25.99 \\
        \small 4 choose 2 & 7.47 & 5.10 & 26.79 \\
        \small \textbf{2 choose 2 (Ours)} & \textbf{7.19} & \textbf{4.89} & \textbf{25.94} \\
        \small 2 choose 1 & 7.21 & \textbf{4.89} & 26.03 \\
        \small 1 choose 1 & 7.23 & 4.90 & 26.14 \\
        \bottomrule
    \end{tabular*}
    \end{small}
    \vspace{-12pt}
\end{table}

\subsection{Ablation Studies and Analysis}


To investigate the impact of MoE adaptor design, we conduct ablation experiments under the full 2,200-hour multi-dataset training setup. Table~\ref{tabel2} compares four configurations with different numbers of experts and top-$k$ selection strategies.

The results indicate that the expert configuration influences performance. To ensure a fair comparison, we adjust the hidden dimensions of each variant to maintain comparable parameter counts. The single-expert configuration (1 choose 1) achieves a MER of 7.23\%, which is inferior to the proposed multi-expert setting. Among the MoE variants, the proposed 2 choose 2 configuration achieves the best overall performance (7.19\% MER), slightly outperforming both 4 choose 2 (7.47\% MER) and 2 choose 1 (7.21\% MER), while the fully activated 4 choose 4 setting yields 7.33\% MER. These findings suggest that dense top-2 activation is marginally more beneficial than sparse top-1 selection, while increasing the number of experts from 2 to 4 does not yield additional gains.

We hypothesize that this behavior is related to the bilingual nature of the code-switching task. Since only Mandarin and English are involved, two experts may be sufficient to capture their distinct acoustic-phonetic characteristics. From a capacity perspective, the MoE adaptor provides scalable modeling flexibility that enables expert specialization without proportionally increasing computation; however, in this bilingual setting, additional experts may introduce unnecessary complexity and optimization difficulty without clear representational benefits.



\begin{figure}[t!]
	\centering
	\includegraphics[width=\linewidth]{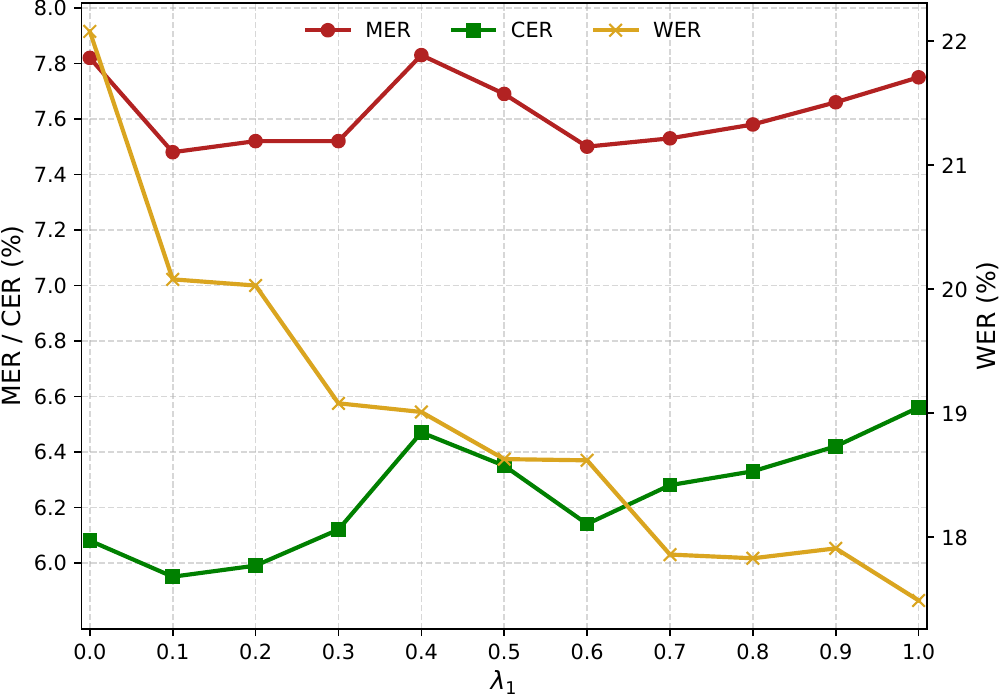}
    \vspace{-16pt}
    \caption{Performance of backbone + hotword prompt models with different $\lambda_1$ using MER, CER and WER.}
	\label{fig2}
    \vspace{-20pt}
\end{figure}

We further analyze the impact of the bias loss weight $\lambda_1$ in Equation~\ref{eq6} on model performance. This ablation is conducted on the ASRU2019-CS training set using the backbone + hotword prompt configuration, as the bias loss is directly associated with the hotword module. Figure~\ref{fig2} presents MER, CER, and WER across $\lambda_1$ values ranging from 0.0 to 1.0.


The results reveal a trade-off between overall MER and the balance between CER and WER. The lowest MER (7.48\%) is achieved at $\lambda_1 = 0.1$, indicating that a relatively small bias weight benefits overall recognition accuracy. As $\lambda_1$ increases, CER and WER become more balanced. At $\lambda_1 = 0.6$, the MER is 7.50\%, only slightly higher than the optimum, while CER and WER are more evenly distributed (6.14\% and 18.62\%, respectively) compared to the $\lambda_1 = 0.1$ setting (5.95\% and 20.08\%).

This behavior reflects the role of the bias loss: increasing its weight strengthens hotword supervision, slightly degrading CER while improving WER on English tokens. Based on this trade-off, we adopt $\lambda_1 = 0.6$ in the full LLM-HB model to achieve a more balanced performance across the two languages.

\section{Conclusion}
In this work, we presented LLM-HB, a framework designed to enhance code-switching ASR through LLM-guided hotword biasing. To the best of our knowledge, this is the first attempt to incorporate LLM-guided biasing into code-switching ASR. The core components include a MoE adaptor that dynamically integrates speech and text representations, an auxiliary language head that provides language identity supervision, and a hotword prompt that introduces targeted biasing signals. Experimental results on ASRU2019-CS demonstrate that the complete model achieves substantial improvements over the backbone, reducing MER from 7.34\% to 5.85\%. Ablation studies further show that dense top-2 expert activation is more effective than sparse alternatives in bilingual settings, and that the bias loss weight regulates the trade-off between character-level and word-level accuracy. These findings confirm the effectiveness of LLM-guided biasing for code-switching ASR and provide insights for future multilingual speech research. In future work, we plan to extend the approach to address near-homophones and to scale to larger hotword vocabularies.




\section{Acknowledgments}
\vspace{-1mm}

This work was supported in part by China NSFC project under Grants No. U25A20409, and in part by SJTU Med-X (Medicine \& Engineering) Translational Research Grant (YG2025LC09).
\vspace{-1mm}

\section{Generative AI Use Disclosure}
\vspace{-1mm}
Artificial intelligence (AI) tools were used exclusively for language polishing and did not contribute to the scientific content of this paper.
\vspace{-1mm}

\bibliographystyle{IEEEtran}
\vspace{-1mm}
\bibliography{mybib}

@INPROCEEDINGS{Povey_ASRU2011_2011,
author = {Povey, Daniel and Ghoshal, Arnab and Boulianne, Gilles and Burget, Lukas and Glembek, Ondrej and Goel, Nagendra and Hannemann, Mirko and Motlicek, Petr and Qian, Yanmin and Schwarz, Petr and Silovsky, Jan and Stemmer, Georg and Vesely, Karel},
title = {The Kaldi Speech Recognition Toolkit},
booktitle = {Proc. IEEE 2011 Workshop on Automatic Speech Recognition and Understanding},
year = {2011},
}

@inproceedings{gulati20_interspeech,
  title     = {{Conformer: Convolution-augmented Transformer for Speech Recognition}},
  author    = {Anmol Gulati and James Qin and Chung-Cheng Chiu and Niki Parmar and Yu Zhang and Jiahui Yu and Wei Han and Shibo Wang and Zhengdong Zhang and Yonghui Wu and Ruoming Pang},
  year      = {2020},
  booktitle = {Proc. Interspeech},
  pages     = {5036--5040},
}

@ARTICLE{8068205,
  author={Watanabe, Shinji and Hori, Takaaki and Kim, Suyoun and Hershey, John R. and Hayashi, Tomoki},
  journal={IEEE Journal of Selected Topics in Signal Processing}, 
  title={Hybrid CTC/Attention Architecture for End-to-End Speech Recognition}, 
  year={2017},
  volume={11},
  number={8},
  pages={1240-1253},
}

@INPROCEEDINGS{6289015,
  author={Vu, Ngoc Thang and Lyu, Dau-Cheng and Weiner, Jochen and Telaar, Dominic and Schlippe, Tim and Blaicher, Fabian and Chng, Eng-Siong and Schultz, Tanja and Li, Haizhou},
  booktitle={Proc. IEEE International Conference on Acoustics, Speech and Signal Processing (ICASSP)}, 
  title={A first speech recognition system for Mandarin-English code-switch conversational speech}, 
  year={2012},
  pages={4889-4892},
}

@inproceedings{liu21d_interspeech,
  title     = {{End-to-End Language Diarization for Bilingual Code-Switching Speech}},
  author    = {Hexin Liu and Leibny Paola García Perera and Xinyi Zhang and Justin Dauwels and Andy W.H. Khong and Sanjeev Khudanpur and Suzy J. Styles},
  year      = {2021},
  booktitle = {Proc. Interspeech},
  pages     = {1489--1493},
}

@inproceedings{lu20f_interspeech,
  title     = {{Bi-Encoder Transformer Network for Mandarin-English Code-Switching Speech Recognition Using Mixture of Experts}},
  author    = {Yizhou Lu and Mingkun Huang and Hao Li and Jiaqi Guo and Yanmin Qian},
  year      = {2020},
  booktitle = {Proc. Interspeech},
  pages     = {4766--4770},
}

@inproceedings{NIPS2017_3f5ee243,
 author = {Vaswani, Ashish and Shazeer, Noam and Parmar, Niki and Uszkoreit, Jakob and Jones, Llion and Gomez, Aidan N and Kaiser, \L ukasz and Polosukhin, Illia},
 booktitle = {Proc. Advances in Neural Information Processing Systems},
 pages = {5598--6008},
 title = {Attention is All you Need},
 volume = {30},
 year = {2017}
}

@INPROCEEDINGS{9747537,
  author={Yan, Brian and Zhang, Chunlei and Yu, Meng and Zhang, Shi-Xiong and Dalmia, Siddharth and Berrebbi, Dan and Weng, Chao and Watanabe, Shinji and Yu, Dong},
  booktitle={Proc. IEEE International Conference on Acoustics, Speech and Signal Processing (ICASSP)}, 
  title={Joint Modeling of Code-Switched and Monolingual ASR via Conditional Factorization}, 
  year={2022},
  pages={6412-6416},
}

@inproceedings{tian22c_interspeech,
  title     = {{LAE: Language-Aware Encoder for Monolingual and Multilingual ASR}},
  author    = {Jinchuan Tian and Jianwei Yu and Chunlei Zhang and Yuexian Zou and Dong Yu},
  year      = {2022},
  booktitle = {Proc. Interspeech},
  pages     = {3178--3182},
}

@INPROCEEDINGS{10389798,
  author={Chen, Peikun and Yu, Fan and Liang, Yuhao and Xue, Hongfei and Wan, Xucheng and Zheng, Naijun and Zhou, Huan and Xie, Lei},
  booktitle={Proc. IEEE Automatic Speech Recognition and Understanding Workshop (ASRU)}, 
  title={{BA-MoE: Boundary-Aware Mixture-of-Experts Adapter for Code-Switching Speech Recognition}}, 
  year={2023},
  pages={1-7},
}

@inproceedings{DBLP:conf/iclr/ShazeerMMDLHD17,
  author       = {Noam Shazeer and
                  Azalia Mirhoseini and
                  Krzysztof Maziarz and
                  Andy Davis and
                  Quoc V. Le and
                  Geoffrey E. Hinton and
                  Jeff Dean},
  title        = {Outrageously Large Neural Networks: The Sparsely-Gated Mixture-of-Experts
                  Layer},
  booktitle    = {Proc. 5th International Conference on Learning Representations, {ICLR}},
  year         = {2017},
}

@inproceedings{20260056940,
title = {MoE Adapter for Large Audio Language Models: Sparsity, Disentanglement, and Gradient-Conflict-Free},
journal = {arXiv},
author = {Lei, Yishu and He, Shuwei and Hu, Jing and Zhang, Dan and Luo, Xianlong and Zhu, Danxiang and Feng, Shikun and Liu, Rui and He, Jingzhou and Sun, Yu and Wu, Hua and Wang, Haifeng},
year = {2026},
booktitle = {{arXiv preprint arXiv:2601.02967}},
}

@INPROCEEDINGS{10095878,
  author={Liu, Hexin and Xu, Haihua and Garcia, Leibny Paola and Khong, Andy W. H. and He, Yi and Khudanpur, Sanjeev},
  booktitle={Proc. IEEE International Conference on Acoustics, Speech and Signal Processing (ICASSP)}, 
  title={Reducing Language Confusion for Code-Switching Speech Recognition with Token-Level Language Diarization}, 
  year={2023},
  pages={1-5},
}

@INPROCEEDINGS{10448335,
  author={Liu, Hexin and Garcia, Leibny Paola and Zhang, Xiangyu and Khong, Andy W. H. and Khudanpur, Sanjeev},
  booktitle={Proc. IEEE International Conference on Acoustics, Speech and Signal Processing (ICASSP)}, 
  title={Enhancing Code-Switching Speech Recognition With Interactive Language Biases}, 
  year={2024},
  pages={10886-10890},
}

@INPROCEEDINGS{10096317,
  author={Yu, Haibin and Hu, Yuxuan and Qian, Yao and Jin, Ma and Liu, Linquan and Liu, Shujie and Shi, Yu and Qian, Yanmin and Lin, Edward and Zeng, Michael},
  booktitle={Proc. IEEE International Conference on Acoustics, Speech and Signal Processing (ICASSP)}, 
  title={Code-Switching Text Generation and Injection in Mandarin-English ASR}, 
  year={2023},
  pages={1-5},
}

@inproceedings{shakeel24_interspeech,
  title     = {{Contextualized End-to-end Automatic Speech Recognition with Intermediate Biasing Loss}},
  author    = {Muhammad Shakeel and Yui Sudo and Yifan Peng and Shinji Watanabe},
  year      = {2024},
  booktitle = {Proc. Interspeech},
  pages     = {3909--3913},
}

@inproceedings{sudo25_interspeech,
  title     = {{DYNAC: Dynamic Vocabulary-based Non-Autoregressive Contextualization for Speech Recognition}},
  author    = {Yui Sudo and Yosuke Fukumoto and Muhammad Shakeel and Yifan Peng and Chyi-Jiunn Lin and Shinji Watanabe},
  year      = {2025},
  booktitle = {Proc. Interspeech},
  pages     = {2215--2219},
}

@INPROCEEDINGS{9415054,
  author={Han, Minglun and Dong, Linhao and Zhou, Shiyu and Xu, Bo},
  booktitle={Proc. IEEE International Conference on Acoustics, Speech and Signal Processing (ICASSP)}, 
  title={Cif-Based Collaborative Decoding for End-to-End Contextual Speech Recognition}, 
  year={2021},
  pages={6528-6532},
}

@INPROCEEDINGS{10446106,
  author={Shi, Xian and Yang, Yexin and Li, Zerui and Chen, Yanni and Gao, Zhifu and Zhang, Shiliang},
  booktitle={ICASSP 2024 - 2024 IEEE International Conference on Acoustics, Speech and Signal Processing (ICASSP)}, 
  title={SeACo-Paraformer: A Non-Autoregressive ASR System with Flexible and Effective Hotword Customization Ability}, 
  year={2024},
  pages={10346-10350},
}

@INPROCEEDINGS{10832154,
  author={Yang, Guanrou and Ma, Ziyang and Gao, Zhifu and Zhang, Shiliang and Chen, Xie},
  booktitle={Proc. IEEE Spoken Language Technology Workshop (SLT)}, 
  title={CTC-Assisted LLM-Based Contextual ASR}, 
  year={2024},
  pages={126-131},
}

@ARTICLE{8733051,
  author={Johnson, Jeff and Douze, Matthijs and Jégou, Hervé},
  journal={IEEE Transactions on Big Data}, 
  title={Billion-Scale Similarity Search with GPUs}, 
  year={2021},
  volume={7},
  number={3},
  pages={535-547},
}

@inproceedings{gong25_interspeech,
  title     = {{BR-ASR: Efficient and Scalable Bias Retrieval Framework for Contextual Biasing ASR in Speech LLM}},
  author    = {Xun Gong and Anqi Lv and Wangyou Zhang and Zhiming Wang and Huijia Zhu and Yanmin Qian},
  year      = {2025},
  booktitle = {Proc. Interspeech},
  pages     = {4043--4047},
}

@inproceedings{kong25_interspeech,
  title     = {{GLCLAP: A Novel Contrastive Learning Pre-trained Model for Contextual Biasing in ASR}},
  author    = {Yuxiang Kong and Fan Cui and Liyong Guo and Heinrich Dinkel and Lichun Fan and Junbo Zhang and Jian Luan},
  year      = {2025},
  booktitle = {Proc. Interspeech 2025},
  pages     = {5173--5177},
}

@inproceedings{wan2024xcbeffectivecontextualbiasing,
      title={{XCB: an effective contextual biasing approach to bias cross-lingual phrases in speech recognition}}, 
      author={Xucheng Wan and Naijun Zheng and Kai Liu and Huan Zhou},
      year={2024},
      booktitle = {{arXiv preprint arXiv:2408.10524}},
}

@inproceedings{shi2020asru2019mandarinenglishcodeswitching,
      title={{The ASRU 2019 Mandarin-English Code-Switching Speech Recognition Challenge: Open Datasets, Tracks, Methods and Results}}, 
      author={Xian Shi and Qiangze Feng and Lei Xie},
      year={2020},
      booktitle = {{arXiv preprint arXiv:2007.05916}},
}

@InProceedings{pmlr-v202-radford23a,
  title = 	 {Robust Speech Recognition via Large-Scale Weak Supervision},
  author =       {Radford, Alec and Kim, Jong Wook and Xu, Tao and Brockman, Greg and Mcleavey, Christine and Sutskever, Ilya},
  booktitle = 	 {Proc. Proceedings of the 40th International Conference on Machine Learning},
  pages = 	 {28492--28518},
  year = 	 {2023},
  volume = 	 {202},
}

@inproceedings{DBLP:conf/iclr/HuSWALWWC22,
  author       = {Edward J. Hu and
                  Yelong Shen and
                  Phillip Wallis and
                  Zeyuan Allen{-}Zhu and
                  Yuanzhi Li and
                  Shean Wang and
                  Lu Wang and
                  Weizhu Chen},
  title        = {LoRA: Low-Rank Adaptation of Large Language Models},
  booktitle    = {Proc. The Tenth International Conference on Learning Representations, ICLR},
  year         = {2022},
}

@inproceedings{du2018aishell2transformingmandarinasr,
      title={AISHELL-2: Transforming Mandarin ASR Research Into Industrial Scale}, 
      author={Jiayu Du and Xingyu Na and Xuechen Liu and Hui Bu},
      year={2018},
      booktitle = {{arXiv preprint arXiv:1808.10583}},
}

@INPROCEEDINGS{7178964,
  author={Panayotov, Vassil and Chen, Guoguo and Povey, Daniel and Khudanpur, Sanjeev},
  booktitle={Proc. IEEE International Conference on Acoustics, Speech and Signal Processing (ICASSP)}, 
  title={Librispeech: An ASR corpus based on public domain audio books}, 
  year={2015},
  pages={5206-5210},
}

@inproceedings{he-choi-2021-stem,
    title = "The Stem Cell Hypothesis: Dilemma behind Multi-Task Learning with Transformer Encoders",
    author = "He, Han and Choi, Jinho D.",
    booktitle = "Proc. Proceedings of the 2021 Conference on Empirical Methods in Natural Language Processing",
    month = nov,
    year = "2021",
    pages = "5555--5577",
}

@inproceedings{le21_interspeech,
  title     = {{Contextualized Streaming End-to-End Speech Recognition with Trie-Based Deep Biasing and Shallow Fusion}},
  author    = {Duc Le and Mahaveer Jain and Gil Keren and Suyoun Kim and Yangyang Shi and Jay Mahadeokar and Julian Chan and Yuan Shangguan and Christian Fuegen and Ozlem Kalinli and Yatharth Saraf and Michael L. Seltzer},
  year      = {2021},
  booktitle = {Proc. Interspeech},
  pages     = {1772--1776},
}

\end{document}